\documentclass[twocolumn,showpacs,showkeys,aps,prd,nofootinbib,groupedaddress,reprint]{revtex4-1}
\usepackage{amsmath,amssymb,amsfonts,dsfont,mathrsfs,amsthm}
\usepackage{graphicx}
\usepackage{centernot}
\usepackage{hyperref}
\hypersetup{linktocpage,colorlinks=true,urlcolor=blue,linkcolor=blue,citecolor=red}
\usepackage{siunitx}
\usepackage{braket}

\newcommand{\form}[1]{{\boldsymbol{#1}}}

\newcommand{\Agf}{\boldsymbol{\mathcal{A}}}

\newcommand{\cdf}[1][]{\,{\boldsymbol{\mathcal{D}}}{#1}\!}

\newcommand{\D}{\mathscr{D}}
\newcommand{\df}[1][]{\,{\mathbf{d}}{#1}\!}

\newcommand{\F}{\,\boldsymbol{\mathcal{F}}}

\newcommand{\gf}{\,\boldsymbol{\gamma}}

\DeclareMathOperator{\hs}{\,\star\!\!}

\newcommand{\Ri}{\mathcal{R}}

\newcommand{\tor}{\mathcal{T}}

\newcommand{\w}{{\scriptstyle\wedge}\!}

\newcommand{\bps}{\ensuremath{\bar{\psi}}}

\newcommand{\vph}{\ensuremath{\varphi}}

\newcommand\vi[2]{e^{{#1}}_{{#2}}}

\newcommand\vif[1]{\,{\mathbf{e}}^{{#1}}}

\newcommand\gam[1]{\gamma^{{#1}}}

\newcommand\spi[1]{\omega_{{#1}}}

\newcommand\spif[2]{\,{\boldsymbol{\omega}}^{{#1}}{}_{{#2}}}
\newcommand\tspif[2]{\,{\tilde{\boldsymbol{\omega}}}^{{#1}}{}_{{#2}}}

\newcommand{\Rif}[2]{\,\boldsymbol{\mathcal{R}}^{{#1}}{}_{{#2}}}
\newcommand{\tRif}[2]{\,\tilde{\boldsymbol{\mathcal{R}}}^{{#1}}{}_{{#2}}}
\newcommand{\Tf}[1]{\,\boldsymbol{\mathcal{T}}^{#1}}

\newcommand{\contf}[2]{\,\boldsymbol{\mathcal{K}}^{#1}{}_{#2}}

\newcommand{\comm}[2]{\left[#1,#2\right]}

\newcommand{\vev}[1]{\ensuremath{\left<#1\right>}}

\newcommand*{\diag}{\operatorname{diag}}

\newcommand{\Tr}{\operatorname{Tr}}

\newcommand{\beq}{\begin{equation}}
\newcommand{\eeq}{\end{equation}}
\newcommand{\ber}{\begin{eqnarray}}
\newcommand{\eer}{\end{eqnarray}}

\renewcommand{\(}{\left(}
\renewcommand{\)}{\right)}
\renewcommand{\[}{\left[}
  \renewcommand{\]}{\right]}

\newcommand\TD{torsion-descended }
\newcommand\KR{Kalb-Ramond }
\newcommand\BI{Barbero-Immirzi }
\newcommand\UTFSM{Departamento de F\'\i sica, Universidad T\'{e}cnica Federico Santa Mar\'\i a}
\newcommand\CCTVal{Centro Cient\'\i fico Tecnol\'ogico de Valpara\'\i so, \\ Casilla 110-V, Valpara\'\i so, Chile}

\begin{document}

\title{Axions in gravity with torsion}

\author{Oscar \surname{Castillo-Felisola}}
\author{Crist\'obal \surname{Corral}}
\author{Sergey \surname{Kovalenko}}
\author{Iv\'an \surname{Schmidt}}
\affiliation{\UTFSM}
\affiliation{\CCTVal}

\author{Valery~E. \surname{Lyubovitskij}}
%\email{valeri.lyubovitskij@uni-tuebingen.de}  
\affiliation{Institut f\"ur Theoretische Physik, 
  Universit\"at T\"ubingen, 
  Kepler Center for Astro- and Particle Physics, \\ 
  Auf der Morgenstelle 14, D-72076 T\"ubingen, Germany,}
\affiliation{Department of Physics, Tomsk State University, 634050
  Tomsk, Russia,}
\affiliation{Mathematical Physics Department, 
  Tomsk Polytechnic University, Lenin Avenue 30, 
  634050 Tomsk, Russia}

%\email{ivan.schmidt@usm.cl}  

\begin{abstract}
We study a scenario allowing a solution of the strong charge parity problem via the Peccei-Quinn mechanism, implemented in gravity with torsion. In this framework there appears a torsion-related pseudoscalar field known as \KR axion. We compare it with the so-called \BI axion recently proposed in the literature also in the context of the gravity with torsion. We show that they are equivalent from the viewpoint of the effective theory.  The phenomenology of these torsion-descended axions is completely determined by the Planck scale without any additional model parameters. These axions are very light and very weakly interacting with ordinary matter. We briefly comment on their astrophysical and cosmological implications in view of the recent BICEP2 and Planck data. 
\end{abstract}


\maketitle

\section{\label{sec:Introduction}Introduction}

The recent discovery of the Higgs boson at the LHC has completed the list of the known particles, providing the last missing element necessary for the standard model (SM) to be the framework for particle physics. However, it is well known that the SM suffers from various internal problems indicating that this is not a fundamental theory, and in fact it should be considered just as an effective low-energy theory. The strong {\it CP} problem is one of these problems. It emerges from adding to the QCD Lagrangian the so called $\theta$ term
\begin{equation}
  \label{Theta-term-def}
  \form{\mathcal{L}} \supset \,\theta \, \frac{\alpha_{s}}{2\pi} \, \Tr\left(\form{G}\w\,\form{G}\right)\,,
\end{equation}
written in terms of the QCD gluon field strength 2-form $\form{G}$. This is a renormalizable and gauge invariant term, which violates {\it CP} and is allowed in any generic gauge theory in four dimensions. In the SM it contributes to {\it CP}-odd observables such as the neutron electric dipole moment, which is stringently constrained by experiment, pushing the $\theta$ parameter down to $\num{e-10}$.  Since the natural value of this parameter should be of order one, this becomes a fine-tuning problem. The question of why it turns out to be so small is the strong CP-problem. 

A solution of the strong {\it CP} problem has been found by Peccei and Quinn in the periodicity of the nonperturbative QCD $\theta$ vacuum~\cite{Peccei:1977hh,*Peccei:1977ur} by promoting the $\theta$ parameter in Eq.~\eqref{Theta-term-def} to be a field $\theta(x)$. Then the interaction $\theta(x)\Tr\left(\mathbf{G}\w\, \mathbf{G}\right)$ generates in the $\theta$ vacuum a nontrivial potential for $\theta(x)$, selecting a zero vacuum expectation value $\vev{\theta} = 0$. The fluctuations around this vacuum represent a pseudoscalar field $a(x)$, dubbed the axion. Then dynamically the {\it CP}-violating term~\eqref{Theta-term-def} is replaced by the {\it CP}-conserving  interaction $a(x) \Tr\left(\mathbf{G}\w\, \mathbf{G}\right)$. 

The $\theta$ parameter can be promoted to be a field, by means of a pseudoscalar field, $\phi(x)$, of any origin, coupled to the Pontryagin density  $\Tr\left(\mathbf{G}\w\, \mathbf{G}\right)$ of the gluon field. This could be a Goldstone boson of a $U(1)_A$ symmetry, spontaneously broken at some scale much larger than the electroweak scale of $\SI{250}{\GeV}$, to be compatible with the experimental data as well as with astrophysics and cosmology. There are many symmetry based proposals of this kind in the literature, as possible solutions of the strong {\it CP} problem (for a recent review see Ref.~\cite{Peccei:2006as}). A characteristic feature of this approach is that all the couplings of the axion are determined by the scale of symmetry breaking, which is a free parameter.

On the other hand it is well known that various scenarios for the Planckian physics involve axionlike fields~\cite{Ibanez:2012zz,Witten:1984dg,Svrcek:2006yi,Conlon:2006tq}. Those fields can play the same role as the conventional Goldstone-type axions in the solution of the strong {\it CP} problem, but with all their couplings completely determined by the Planck scale. 

In particular the axionlike fields may appear rather naturally in a field theory on the torsionful manifolds with its metric sector treated as a ``rigid'' background. The first scenario of this kind was proposed in Ref.~\cite{Duncan:1992vz}, where an axionlike field appears as a consequence of the constraint imposed on the quantum theory requiring the conservation of the torsion charge, as suggested by the classical theory. 

Recently, in Ref.~\cite{Mercuri:2009zi}, the axion has been introduced as a pseudoscalar field, the so-called \BI (BI) axion, interacting with gravity via the Nieh-Yan (NY) density~\cite{Nieh:1981ww,Plebanski:1977zz,*PhysRevD.22.1915,*Nelson:1980ph}. One of the motivations for the introduction of this field was the possibility of eliminating the confusing divergence present in the $U(1)_A$ rotated fermion measure of the Euclidean path integral on the manifolds with torsion. In addition to the usual Pontryagin density, in this case there appears a Nieh-Yan term, which becomes divergent when the regularization 
is removed~\cite{Chandia:1997hu}. The significance of this divergence was debated in the literature~\cite{Kreimer:1999yp,Chandia:1999az} and a consensus on its status has not yet been reached. In the model of Ref.~\cite{Mercuri:2009zi} this divergence can be absorbed by a redefinition of the \BI field. This axionlike field was also proposed in Ref.~\cite{Lattanzi:2009mg}, in order to solve the strong {\it CP} problem in the Peccei-Quinn spirit. 

In the present paper, we show that the conservation of the torsion charge, within the framework of Ref.~\cite{Duncan:1992vz}, is equivalent to demanding a vanishing Nieh-Yan density. This constraint can be implemented into the quantum theory by means of a Lagrange multiplier, identified with the so-called the \KR (KR) axion~\cite{Kalb:1974yc}, due to its similarity with the axionlike field coming from string theory. 

Despite the starting points of Refs.~\cite{Duncan:1992vz}-\cite{Mercuri:2009zi} seeming to be different, they have the same physical properties when the torsion is integrated out. Therefore, within the effective theory, the \KR~\cite{Kalb:1974yc} and the \BI~\cite{Mercuri:2009zi} axions are equivalent. We rigorously demonstrate this equivalence and study the solution of the  strong {\it CP} problem based on these torsion-descended (TD) axions. Then we examine their possible 
%phenomenological impact as well as  
cosmological and astrophysical implications.

In the present manuscript we concentrate on the discussion of axions in Einstein-Cartan theory of gravity with torsion. For discussions on the role of axions motivated by Chern-Simons-type terms, see Ref.~\cite{Mielke:2006zp}, where as a cosmological application, the accelerated expansion of the Universe has been considered~\cite{Minkevich:2010ru,Mielke:2006zp}.

The article is organized as follows. In Sec.~\ref{sec:gravity} we sketch the classical Einstein-Cartan gravity with torsion. In Sec.~\ref{sec:axion} we discuss the two procedures for quantizing the model, and show that despite their different origin, they are equivalent as effective theories. In Sec.~\ref{sec:strongCP} we consider some cosmological and astrophysical implications of these \TD axions. Finally, we summarize our present study in Sec.~\ref{Conclusions}.

\section{\label{sec:gravity}Classical Gravity Setup}  

We consider the Einstein-Cartan theory of gravity -- a minimal construction of gravitational theory allowing the connection to possess a nonvanishing torsion~\cite{Cartan1923,*Cartan1924,*Cartan1925,RevModPhys.48.393,*Shapiro:2001rz,*Hammond:2002rm}. The Cartan's structure equations, relating curvature and torsion with the vierbein $\vi{a}{\mu}$ and the spin connection $\spi{\mu}{}^{ab}$, read  
\begin{align}
  \df\vif{a} + \spif{a}{b}\,\w\,\vif{b} &= \Tf{a},
  \label{first-str-eq} \\
  \df\spif{a}{c} + \spif{a}{b}\,\w\,\spif{b}{c} &= \Rif{a}{c}, 
  \label{sec-str-eq}          
\end{align} 
where $\vif{a} = e^a_\mu\,dx^\mu$ and $\spif{a}{b} = \left(\omega_\mu\right)^{a}{}_{b}\,dx^\mu$ are the vierbein and spin connection $1$-form respectively. Hereon, bold symbols will denote differential forms, while greek and latin indices stand for spacetime and Lorentz indices respectively. Additionally, the vierbein is related with the curved metric $g_{\mu\nu}$ through
\begin{equation}
  g_{\mu\nu} = \eta_{ab}\,e^{a}_\mu\,e^{b}_\nu\,,
\end{equation}
where $\eta_{ab} = \diag\left(-1,+1,+1,+1\right)$ is the Minkowski metric in four dimensions. The spin connection can be split into \mbox{$\spif{ab}{} = \tspif{ab}{} + \contf{ab}{}$}, where the tilde indicates a torsion-free quantity and the contorsion tensor $ \contf{ab}{}$ encodes the information about the torsion through the relation $\Tf{a} = \contf{a}{b}\w\vif{b}$. Similarly, Eq.~\eqref{sec-str-eq} together with the decomposition of the spin connection yields to
\begin{equation}
  \Rif{ac}{} = \tRif{ac}{} + \tilde{\form{\mathcal{D}}}\contf{ac}{} 
  + \contf{a}{b}\w\contf{bc}{}\,. 
\end{equation}
The Einstein-Cartan action can be written as
\begin{equation}
  \mathcal{S}_{\text{gr}} = \frac{1}{4\kappa^2}
  \int \epsilon_{abcd}\Rif{ab}\w\vif{c}\w\vif{d}\,,
\end{equation}
where $\kappa^2 = 8\pi G_N = 8\pi M^{2}_{Pl}$ with $G_N$ and $M_{Pl}$ being the Newton's constant and the Planck scale, respectively.

In the following, the SM fields are assumed to live in a curved torsionful spacetime. The nontrivial coupling of matter with torsion enters in the fermionic sector through the covariant derivative,
\begin{equation}
  \cdf[\psi] = \df[\psi] + \frac{1}{4}\spif{ab}{}\gamma_{ab}\psi 
  + \imath e \Agf \psi + \imath g \form{B}\psi , 
\end{equation}
where $\Agf$ and $\form{B}$ denote the $U(1)_{em}$ and $SU(3)_c$ gauge boson $1$-forms respectively and $\gamma_{ab}=\tfrac{1}{2}\comm{\gamma_a}{\gamma_b}$. Therefore, the complete model is described by the action
\begin{align}
  \nonumber {\cal S} &= \mathcal{S}_{\text{gr}} - \frac{1}{2}\sum_f 
  \int \(\bps_f \gf\w\hs\cdf[\psi]_f - \cdf[\bps]_f \w\hs\gf \psi_f\) \\
  & \quad - \frac{1}{2}\int\F\w\hs\F 
  - \int\Tr\[\form{G}\w\hs\,\form{G}\]\,, 
  \label{action}
\end{align}
where $\bps = -i\psi^\dagger\gamma^0$ is the usual Dirac adjoint, $\gf = \gamma_a\vif{a}$ and the subscript $f$ indicates the SM fermionic flavors.  
The symbol $\star$ denotes Hodge duality, while $\form{G}$ and $\F$ are the $SU(3)_{c}$ and $U(1)_{em}$ gauge field strength $2$-form respectively.

When the action in Eq.~\eqref{action} is varied with respect to the vierbein field, one obtains the corresponding Einstein-Cartan equation of motion,
\begin{equation}
  \label{eceq}
  \Ri_{ab} -\frac{1}{2}\eta_{ab}\Ri = \kappa^2 \tau_{ab}, 
\end{equation}
where $ \tau_{ab}$ is the energy-momentum tensor of the system. In its form Eq.~\eqref{eceq} looks similar to the Einstein equations derived within general relativity (GR). However, the two equations are different because the presence of torsion gives rise to an antisymmetric part in both sides of the equation, absent in GR.  

On the other hand, the variation of the action with respect to the spin connection yields to an algebraic equation of motion 
for the torsion,
\begin{equation}
  \label{EOM-T}
  \tor_{abc} = -\frac{\kappa^2}{2}\epsilon_{abcd}\sum_f J_f^{5\, d}\,,
\end{equation}
where $J_f^{5\, d} = i \bps_f\gam{d}\gam{5} \psi_f$ is the fermionic axial current. Notice that only the completely antisymmetric part of the torsion couples to the fermionic fields. This part corresponds to its axial irreducible component, 
\begin{equation}
  \label{Af-def} 
  \mathbf{S} %= \hs\Tf{}
  = \tfrac{1}{3!}\epsilon_{abcd}\tor^{abc}\vif{d} \,.
\end{equation}

Now we rewrite  Eq.~\eqref{action}, showing explicitly only those terms which depend on the torsion axial component
\begin{equation}
  \mathcal{S} = \mathcal{S}_0  
  + \frac{3}{4\kappa^2}\int \mathbf{S}\w\hs\,\mathbf{S}
  - \frac{3}{4} \int \mathbf{J}^5 \w\hs\,  \mathbf{S},
  \label{new-action} 
\end{equation}
where $\mathcal{S}_0 = \tilde{\mathcal{S}}_{\text{gr}} + \tilde{\mathcal{S}}_{\psi} + \mathcal{S}_{\rm gk}$ and $\mathcal{S}_{\text{gk}}$ represents the gauge field kinetic terms [the last two terms in Eq.~\eqref{action}]. The entity $\mathbf{J}^5 = \sum_f (J_f^{5})_a\vif{a}$ denotes the axial current 1-form, where the sum runs for all the fermionic flavors $f$.

\section{Quantum theory and TD Axions}
\label{sec:axion}

The quantization of the model with the action given in Eq.~\eqref{new-action} can be carried out on the basis of the path integral representation for the generating functional. However, at present it is unknown if this procedure is applicable to the quantization of the whole gravity sector (for the current status of this problem see, for instance, Refs.~\cite{Daum:2013fu,*Mielke:2013vma}).

In the scenario studied here, the SM fields lie on a torsionful manifold, whose only quantum gravity effects enter through the torsion, while the metric or Riemmanian curvature remains as classical variable. Quantum torsion seems to be easily treatable because the equation of motion for the torsion~\eqref{EOM-T} is algebraic, showing that the torsion is a nonpropagating field, which can be exactly integrated out from the theory. 

However, as it was observed in Ref.~\cite{Duncan:1992vz}, this treatment of the torsion should be done with caution. It follows from Eqs.(10)-(11), that $\mathbf{S} \propto \mathbf{J}^5_f$. Since the action~\eqref{new-action} is $U(1)_{A}$ symmetric, the N\"other current $\mathbf{J}^5_f $ is conserved at classical level, leading, as follows from the above relation, to the conservation of the torsion charge $Q_{S}= \int\hs\, \mathbf{S}$. 

On the other hand, we know that the fermionic measure of the path integral is not $U(1)_{A}$ invariant. This fact manifests as the anomalous nonconservation of $\mathbf{J}^5$ at the quantum level. As pointed out in Ref.~\cite{Duncan:1992vz}, this must be taken into account before integrating out the torsion, in order to maintain the self consistency of the constructed effective theory. Following Ref.~\cite{Duncan:1992vz}, an effective quantum theory can be constructed through a constraint requiring the conservation of the torsion charge $\mathbf{d\hspace*{-0.8mm}\hs\, S} = 0$. Notice that this is a gauge invariant condition, which is important for the self consistency of the SM sector of the theory. Later we show that this condition eliminates the divergent part of the $U(1)_{A}$ anomaly mentioned in Sec. I and affecting the tractability of the quantum theory in the presence of the torsion.  

The quantum generating functional, with this condition incorporated, takes the form 
\begin{equation}
  \mathcal{Z} = \int\prod_\vph \D\vph\, \D\form{S}\, 
  e^{i {\cal S}[\vph]}\;
\delta\( \mathbf{d\hspace*{-0.8mm}\hs\, S} \)\,,
  \label{Z}
\end{equation}
where  ${\cal S}[\vph]$ is the action given in Eq.~\eqref{new-action} and  $\vph$ denotes all the fields except for $\vi{a}{\mu}$, treated as a rigid background. The argument of the delta in Eq.~\eqref{Z} can be passed to the effective action 
using the integral representation 
\begin{equation}
\delta\( \mathbf{d\hspace*{-0.8mm}\hs\, S} \) = \int \D\phi\;e^{\int i \phi \df\hs\, \mathbf{S}}
  = \int \D\phi\; e^{-\int i \df[\phi]\,\w\hs\, \mathbf{S}}\,.
\end{equation}
This allows us to write
\begin{eqnarray}
\nonumber
\mathcal{Z} &=& \int\prod_\vph \D\vph \,e^{i {\cal S}_{0}}\,\int \D\form{S} \D\phi\,\exp\left\{i\int \left[\frac{3}{4\kappa^2} \mathbf{S}\w\hs\,\mathbf{S} \right.\right.\\
  \label{Z-1}
  &-& \left.\left.\frac{3}{4} \mathbf{J}^5 \w\hs\,  \mathbf{S}-\df[\phi]\,\w\hs\, \mathbf{S}\right]\right \}\,,
\end{eqnarray}
where $\mathbf{J}^5 = \sum_f \mathbf{J}^5_f$. 

Since $\mathbf{S}$ is a nonpropagating field, we can integrate it out in the standard way~\cite{Duncan:1992vz,RevModPhys.48.393,*Shapiro:2001rz} carrying out a variable transformation 
\begin{eqnarray}\label{var-change}
\mathbf{S^{\prime}} = \mathbf{S} - \frac{2}{3}\kappa^2\df[\phi] 
- \frac{1}{2} \kappa^{2} \mathbf{J}^5\,,
\end{eqnarray}
with the Jacobian equal to unity. This new variable appears in the exponent in Eq.~\eqref{Z-1} only in the bilinear combination $\mathbf{S}\w\hs\,\mathbf{S}$ and, therefore, can be exactly integrated out. As a result we get the effective action 
\begin{align}
  \nonumber    {\cal S}_{\text{eff}} &=  {\cal S}_0 
  - \frac{3\kappa^2}{16}
  \int \mathbf{J}^5\w\hs\,\mathbf{J}^5  -\\
  &\quad - \frac{1}{2}\int \df[\Phi]\,\w\hs\df[\Phi] 
  + \sqrt{\frac{3}{2}}\frac{\kappa}{2}\,\int \Phi\df\hs\,\mathbf{J}^5\,.
  \label{Seff}
\end{align}
For convenience we have made a redefinition $\Phi = \sqrt{2/3}\kappa \, \phi$. Notice that the integration out of the torsion makes  $\Phi(x)$ a dynamical field with the canonical kinetic term. As follows from the last term in Eq.~\eqref{Seff} this field is pseudoscalar. It is what we called in the introduction KR axion field.  

At the quantum level, the last term of Eq.~\eqref{Seff} is nothing but the axial anomaly~\cite{PhysRev.177.2426,*Bell:1969ts}. In the path integral language $\df\hs\,\mathbf{J}^5\neq 0$ is the manifestation of the $U(1)_A$ noninvariance of the fermionic measure ~\cite{PhysRevLett.42.1195}, mentioned in the Introduction.

On the Riemann-Cartan manifolds the axial anomaly was first studied in Refs.~\cite{Obukhov:1982da,*Obukhov:1983mm,*Yajima:1985jd}. However, as shown in Ref.~\cite{Chandia:1997hu}, the computation of 
such an anomaly gives rise, in general, to an additional previously missed term, called Nieh-Yan topological density~\cite{Nieh:1981ww} so that under a $U(1)_{A}$ rotation of the fermion fields $\psi$ 
the fermion measure experiences a nontrivial variation~\cite{Obukhov:1982da,*Obukhov:1983mm,*Yajima:1985jd,Chandia:1997hu}
\begin{equation}
  \begin{split}
    \D\psi \D\bar\psi \rightarrow& \D\psi \D\bar\psi \times \exp \bigg\{i \alpha \int \Big[\frac{\alpha_{\text{em}} \bar{Q}^{2}}{\pi} \F\w\F  \\
      &+ \frac{\alpha_s N_{q}}{2\pi}\Tr\[\mathbf G\w\, \mathbf G\] 
      + \frac{N_{f}}{8\pi^2}\Rif{ab}{}\w\Rif{}{ab} \\
      & + 2 M^{2} \left(\boldsymbol{\mathcal{T}}_{a}\w\Tf{a} 
      -  \mathbf{e}_{a} \w \, \mathbf{e}_{b}\,  \w \Rif{ab}{}\right)
      \Big] \bigg\}\, .
  \end{split}
  \label{measure-var}
\end{equation}
Here $\alpha_{\text{em}}$ and $\alpha_{s}$ are the electromagnetic and QCD couplings, respectively,  $N_f$ is the total number of fermionic flavors, $N_{q}$ is the number of quarks and $\bar{Q}^{2} = \sum_{f} Q^{2}_{f}$, where $Q_{f}$ is the charge of $f$ fermionic flavor. The last term in Eq.~(\ref{measure-var})  is NY topological density with the regulator multiplier being divergent when the regularization is removed, $M\rightarrow \infty$. As mentioned in Sec.~\ref{sec:Introduction} the status of this divergence is still debated in the literature. 

However we find that in the approach of  Ref.~\cite{Duncan:1992vz} it is irrelevant since the NY term $ \form{\mathcal{N}}$ vanishes identically due to the condition $\mathbf{d\hspace*{-0.8mm}\hs\, S} = 0$ imposed on quantum theory 
by insertion of the corresponding delta function in Eq.~(\ref{Z}). In fact  this follows from the identity 
derived in Ref.~\cite{Nieh:1981ww},
\begin{equation}
  \label{NY}
  \form{\mathcal{N}} \equiv \Tf{a}\wedge\form{\cal{T}}_{a} - \form{\cal{R}}_{ab}\wedge\vif{a}\wedge\vif{b} = \df\left(\vif{a}\wedge\form{\cal{T}}_{a}\right),
\end{equation}
and the definition of the field $\mathbf{S}$ in Eq. (\ref{Af-def}) written in the form
\begin{eqnarray}\label{def-S-1}
\hs\, \mathbf{S}\propto \vif{}_{a}\w\,\Tf{a}.
\end{eqnarray}
Then from Eqs.~(\ref{NY})-(\ref{def-S-1}) it follows that
\begin{eqnarray}\label{dS-NY-zero}
\mathbf{d\hspace*{-0.8mm}\hs\, S} = 0 \Rightarrow \boldsymbol{\mathcal{N}} = 0.
\end{eqnarray}
Thus, neglecting the Nieh-Yan term in the axial anomaly, we can write for the axial current
\begin{equation}
  \begin{split}
    \df\hs\;\mathbf{J}^5 &= -\frac{\alpha_{\text{em}} 
    \bar{Q}^{2}}{\pi} \F\w\F 
    - \frac{\alpha_s N_{q}}{2\pi}\Tr\[\mathbf G\w\, \mathbf G\] \\
    &\quad - \frac{N_{f}}{8\pi^2}\tRif{ab}{}\w\tRif{}{ab}.
  \end{split}
  \label{chiral-anomaly}
\end{equation}  
The right-hand side is written in terms of torsion-free quantities. This is attainable by the introduction of proper 
counterterms, as shown in Ref.~\cite{Duncan:1992vz}.

Now we substitute the identity~\eqref{chiral-anomaly} into Eq.~\eqref{Seff} and obtain the resulting effective action of the model,
\begin{align}
  \label{totalZ}
    {\cal S}_{\rm eff} &=
    {\cal S}_0 - \frac{1}{2 f_{\Phi}^{2}}\int\!\mathbf{J}^5\w\hs\,\mathbf{J}^5  
    - \frac{\alpha_{\text{em}} \bar{Q}^{2}}{\pi f_\Phi}\int\!\Phi\F\w\,\F\, 
    \notag
    \\
    & - \frac{1}{2}\int\! \df[\Phi]\,\w\hs\df[\Phi]
    - \frac{1}{8\pi^2}\,\int\!\left(\Theta  
    + \frac{N_{f}}{f_{\Phi}} \Phi \right) \tRif{ab}{}\w\tRif{}{ab} \notag
    \\
    & - \frac{\alpha_s}{2\pi}\,\int\!\left(\theta  
    + \frac{N_{q}}{f_{\Phi}}\Phi\right)  \Tr[\mathbf G\w\,\mathbf G].
\end{align}
Here we introduced a parameter  
\begin{equation}
  \label{scale-f}
  f_{\Phi} = \kappa^{-1}\sqrt{8/3} \simeq \SI{4e18}{\GeV},
\end{equation}
analogous to the decay constants of fields with derivative couplings, such as Goldstones of spontaneously broken symmetries. 

In the effective action (\ref{totalZ}) we added the QCD and the gravitational $\theta$ and $\Theta$ terms. They are the gauge and gravitational Pontryagin densities allowed by the gauge symmetries of the theory. These terms are also needed for the model completeness, and play the role of counterterms for the axial anomaly quantum corrections. They do not affect the previous derivation, since due to their topological nature they do not change the equations of motion.

Recently in Ref.~\cite{Mercuri:2009zi} there  has been proposed an alternative scenario in gravity with torsion  also leading to an axionlike field. This scenario is inspired, in particular, by the Chern-Simons modified gravity motivated in its turn by string theory. The gravitational action according to Ref.~\cite{Mercuri:2009zi} is modified at the classical level by the term 
\begin{eqnarray}
\label{BI-mod}
{\cal S}_{\rm tot} = {\cal S}  + \int \beta(x)\, \boldsymbol{\mathcal{N}}\,,
\end{eqnarray}
where ${\cal S}$ is the action given in Eq.~\eqref{action} or~\eqref{new-action}. This action, being used in the quantum generating functional, allows one to absorb the divergent NY part of  the anomalous $U(1)_{A}$ variation of the fermion measure~(\ref{measure-var}) by a renormalization of the field $\beta(x)$ called in Ref.~\cite{Mercuri:2009zi} the BI axion. The field $\beta(x)$ becomes a dynamical field with the canonical kinetic term after excluding a nondynamical torsion field using the classical equation of motion 
\begin{eqnarray}\label{BI-EOM}
\mathbf{S} = \frac{2}{3} \kappa^{2} \df[\beta] + \frac{1}{2} \kappa^{2} \mathbf{J}^5\,,
\end{eqnarray} 
derived from the action~\eqref{BI-mod}. This is equivalent to integrating out the torsion field $\mathbf{S}$ in the generating functional carrying out the transformation~\eqref{var-change}. Note that the field $\beta(x)$ in the classical action~\eqref{BI-mod} is nothing but a Lagrange multiplier setting the classical level constraint $\boldsymbol{\mathcal{N}} = 0$. Now one can immediately realize that in view of the identities~\eqref{NY})-\eqref{def-S-1} it is equivalent to the constraint 
$\mathbf{d\hspace*{-0.8mm}\hs\, S} = 0$, which in the approach of  Ref.~\cite{Duncan:1992vz} was set at quantum level as a constraint incorporated in the generating functional \eqref{Z} or \eqref{Z-1}. On the other hand both approaches lead to the same effective quantum theory with the effective action given in Eq.~(\ref{totalZ}) with the identification of the KR and BI axions $\Phi(x) \equiv \beta(x)$. As we have seen KR and BI axions originate from rather different treatments 
of quantum theory in the presence of the torsionful gravity. Nevertheless from the point of view of low-energy effective theory and the resulting phenomenology they both are equivalent particles, which we call from now on TD axions. 

Additionally, the TD axions may also appear in the context of the torsion-induced quintessential axions~\cite{Mielke:2006zp}. In this framework, the axial current is modified by the addition of the Chern-Simons-type terms, in order to be conserved in the zero mass limit. The complete cancellation of the torsion sector in the anomaly can be addressed requiring the torsion to be an exterior derivative of a pseudoscalar field, identified later with the axion~\cite{Mielke:2006zp}. Remarkably, this approach leads to the same effective theory as Refs.~\cite{Duncan:1992vz,Mercuri:2009zi} for a constant dilaton field.

\section{Phenomenology and cosmology with TD axions} 
\label{sec:strongCP}

In the effective action in Eq.~\eqref{totalZ} of the considered TD axions the last term is the most important in the context of the strong {\it CP} problem. The presence of  the coupling of an axionlike field  to the gluon field Pontryagin density is the necessary and sufficient condition for the solution of the strong {\it CP} problem via the Peccei-Quinn mechanism. The TD axion decay constant  $f_{\Phi}$ introduced in Eq.~\eqref{totalZ} represents a typical energy scale of the model, related to the Planck scale. Thus, the TD axions $\Phi$ emerge without any accompanying free parameter. This drastically distinguishes them from the axions introduced as Goldstone fields of spontaneously broken symmetries, requiring at least one free model parameter, i.e., the scale of symmetry breaking.

In principle in certain models both the TD and Goldstone axions can coexist mixing with each other~\cite{Lattanzi:2009mg}. We do not consider this case here since in the presence of the TD axions, solving the strong {\it CP} problem without free parameters, introduction of other axions looks excessive.  

Focusing on the last term in Eq.~\eqref{totalZ}, let us write down \mbox{$\theta(x) = \theta + \Phi(x)N_{q}/f_{\Phi}$}. The main point of the Peccei-Quinn mechanism is that the coupling \mbox{$\sim\int\theta(x)\Tr\left[\form{G}\w\,\form{G}\right]$} generates a nontrivial potential for the $\theta(x)$ field. The periodicity of this potential in $\theta$~\cite{Jackiw:1976pf} selects the unique nontrivial minimum $\vev{\theta(x)} = 0$,  corresponding to  $\vev{\Phi(x)} = - \theta \, f_{\Phi}/N_q$. Perturbations around this vacuum generate the physical pseudoscalar  axion field $a(x) = \Phi(x) - \vev{\Phi(x)}$. Thus the only surviving piece of the last term of Eq.~\eqref{totalZ} is the {\it CP}-conserving interaction $a(x)\Tr\left[\form{G}\w\,\form{G}\right]$. This solves the strong {\it CP} problem with the help of the TD axions. The mass $m_{a}$ of the TD axion 
can be calculated in the usual way, as is done for any axion field (for a review cf. Ref.~\cite{Peccei:1998jv}), 
and depends only on the parameter $f_{\Phi}$ defined in Eq. (\ref{scale-f}). The nontrivial mass is generated by instantons and
for the value in Eq.~\eqref{scale-f} it turns out to be
\begin{equation}
\label{a-mass}
  m_a \approx m_{\pi}\,\frac{f_{\pi}}{f_{\Phi}} \frac{\sqrt{m_{u} m_{d}}}{m_{u}+ m_{d}} \sim \SI{e-12}{eV},  
\end{equation}
where $f_{\pi} = 93$ MeV is the $\pi$-meson decay constant and $m_{\pi}, m_{u, d}$ are the masses of $\pi$-meson and $u, d$ quarks. Such an extremely light particle, having the inverse Planck mass suppressed interactions with gauge fields, is unobservable in laboratory experiments. 

Nonetheless, an axion with these properties may play a significant cosmological and astrophysical role, since it must satisfy the existing limits related to its origins. These aspects of the \TD axions considered here, have been studied in Ref.~\cite{Lattanzi:2009mg}. It has been shown that such axions safely pass all the known astrophysical constraints, which originate from the energy loss of a stellar core in the form of axions and its impact on stellar evolution.  

An interesting cosmological prediction applied to the \TD axions, also discussed in Ref.~\cite{Lattanzi:2009mg}, is the production of axion isocurvature perturbations in which amplitude is constrained from the above by WMAP data~\cite{Visinelli:2009zm, *Hamann:2009yf}. Assuming that they are the dominant component of dark matter in the Universe it was found for the upper limit~\cite{Lattanzi:2009mg} 
\begin{eqnarray}\label{Hubble-lim}
  H_{I} \leq \SI{E10}{\GeV}\,,
\end{eqnarray}
for the Hubble expansion rate $H_{I}$ during inflation. Now we can estimate the tensor-to-scalar ratio $r = P_{T}/P_{S}$ 
using expressions for the power spectra of the scalar $P_{S}$ and tensor $P_{T}$ perturbations~\cite{Weinberg:2008}
\begin{eqnarray}
  \label{PS-PT}
  P_{S} \approx \frac{1}{8 \pi^{2}} \left(\frac{H_{I}^{2}}{\epsilon M_{P}^{2}} \right),
  \quad
  P_{T} \approx \frac{2}{\pi^{2}} \left(\frac{H_{I}^{2}}{M_{P}^{2}} \right).
\end{eqnarray}
Here $M_{\rm Pl} = 1/\sqrt{8\pi G_{N}}\approx \SI{2.44E18}{\GeV}$ is the reduced Planck scale and $\epsilon$ is the standard 
slow-roll parameter. We use the Planck Collaboration result $P_{S} \approx \num{2.19E-9}$ (see Refs.~\cite{Ade:2013uln,Ho:2014xza}) and find from Eqs.~\eqref{PS-PT}  
\begin{eqnarray}\label{HI-r}
  H_{I} \approx \SI{2.5E14}{\GeV} \sqrt{r}.
\end{eqnarray}
Using the limit~\eqref{Hubble-lim} one finds the prediction for the cosmology with \TD axions
\begin{eqnarray}\label{r-lim}
  r\leq \num{1.6E-9}.
\end{eqnarray}
This is in dramatic contradiction with the recently published result by BICEP2~\cite{Ade:2014xna}, $r=0.2^{+0.07}_{-0.05}$. 
Nevertheless, the situation has recently changed after the publication of the Planck Collaboration detailed analysis of the 
impact of the diffuse galactic dust polarized emission on the measurements of the polarization of the cosmic microwave  background (CMB)~\cite{Adam:2014bub}. It has been shown that the BICEP2 result can be accounted for in the presence of this dust. Thus, the \TD axions, considered in the present paper, are currently not excluded by the cosmological data.  Their cosmological test via the tensor-to-scalar ratio $r$ is postponed for the future. Results of improved measurements of $r$, taking into account the complications with the diffuse galactic dust, are expected to come in the near future from the Keck Array~\cite{Ogburn:2012ma} and BICEP3~\cite{Ahmed:2014ixy} telescopes.  The first results of the joint analysis of  BICEP2/Keck Array and Planck data have been recently issued~\cite{Ade:2015tva} showing only an upper limit $r< 0.12$ at 95\% C.L. New results from \mbox{BICEP3}, which will improve this limit, are expected during 2015 and 2016 seasons \cite{Ade:2015tva}. 

The following final note might be in order. If the considered scenario is incorporated into the extradimensional setup~\cite{ADD1,*RS2} amended with the torsion \cite{Lebedev:2002dp,*Castillo-Felisola:2014xba} the fundamental $D$-dimensional Planck scale $M_*$ could be reduced down to the $\SI{}{\TeV}$ values, dramatically changing the phenomenology and cosmology of the TD axions. In fact, making a rescaling $M_{\rm Pl}\rightarrow M_*\geq \SI{100}{\GeV} $ in Eqs.~\eqref{scale-f} and \eqref{a-mass}, we find the values $m_{a}\leq\SI{38}{\keV}$ and $f_{\Phi}\geq \SI{100}{\GeV}$. Then for the rate of $a\rightarrow \gamma\gamma$ we get $\Gamma_{a\gamma\gamma}\leq \SI{e-16}{\eV}$. As to the cosmological aspects of the extradimensional TD axions, they require a dedicated study. In particular, the value of the tensor-to-scalar ratio $r$ cannot be obtained from~\eqref{r-lim} by the simple rescaling of $M_{\rm Pl}$.  Let us recall that the bound in Eq.~(\ref{r-lim}) was obtained using the limit~\eqref{Hubble-lim} derived in Ref.~\cite{Lattanzi:2009mg} with the assumption that the axions be out of thermal equilibrium with photons during inflation. This condition can be violated for the values of $\Gamma_{a\gamma\gamma}$ given above. The role of the extradimensional TD axions as a dark matter candidate should also be reconsidered. The corresponding study is in progress and its results will be published elsewhere.

\section{\label{Conclusions}Conclusions}

We considered a solution of the strong {\it CP} problem via the Peccei-Quinn mechanism, implemented into the theory of gravity with torsion. We showed that the self-consistency condition of quantum theory $\mathbf{d\hspace*{-0.8mm}\hs\, S} = 0$  proposed in Ref.~\cite{Duncan:1992vz} is equivalent to the requirement of vanishing Nieh-Yan topological density on the spacetime manifold. The Lagrange multiplier field, incorporating this constraint, leads to the \TD axion coupled to the gluon Pontryagin density, $\Tr\left[\form{G}\w\,\form{G}\right]$, and therefore allows application of the Peccei-Quinn mechanism for solving the strong {\it CP} problem.  

We considered the \KR and the \BI axions proposed in the literature from quite different theoretical perspectives. We found that from the viewpoint of the effective theory these two \TD axions are equivalent. 

An important property of  the \TD axions is that their phenomenology has no free parameters, rather they are completely determined by the Planck scale or, equivalently, by Newton's gravity constant. The \TD axion masses and their characteristic decay constants, are extremely small due to the Planck suppression, typical for this family of axions rooted in gravity. We demonstrated the compatibility of the \TD axions with all the existing cosmological and astrophysical limitations, as well as prospects for testing them in the near future measurements of the tensor-to-scalar ratio of the perturbation modes of the CMB. We also estimated the possible role of extra dimensions in phenomenology and cosmology of \TD axions.

%% \medskip 

\begin{acknowledgments}
  We thank A.~Toloza, G.~Moreno, C.~Dib and J.~Zanelli for fruitful discussions. This work was supported by the DFG under Grant No.~LY~114/2-1, by Tomsk State University Competitiveness Improvement Program, by CONICYT (Chile) under Grants No.~21130179, No.~80140097 and No.~79140040 and FONDECYT (Chile) under Grants No.~1150792 and No.~1140390. C.~C. thanks the University of T\"ubingen for hospitality during the completion of this work.
\end{acknowledgments}

\end{document}